\documentclass[acmsmall]{acmart}

\renewcommand\footnotetextcopyrightpermission[1]{} 
\usepackage{listings}

\usepackage{amsmath,amssymb,amsfonts}
\usepackage{graphicx}
\usepackage{xcolor}
\usepackage{subfigure}
\usepackage{booktabs}
\usepackage{multirow}
\usepackage{makecell}
\usepackage{threeparttable}
\usepackage{enumitem} 
\usepackage{url}
\usepackage{bbding}
\usepackage{diagbox}
\usepackage[ruled, linesnumbered]{algorithm2e}
\usepackage[breakable]{tcolorbox}
\usepackage{braket}
\usepackage{longtable}

\AtBeginDocument{%
	\providecommand\BibTeX{{%
			\normalfont B\kern-0.5em{\scshape i\kern-0.25em b}\kern-0.8em\TeX}}}
		
\begin{document}
	
\title{Quantum-Based Software Engineering}

\author{Jianjun Zhao}
\email{zhao@ait.kyushu-u.ac.jp}
\affiliation
{
	\institution{Kyushu University}
	\country{Japan}
}

\begin{abstract}
Quantum computing has demonstrated the potential to solve computationally intensive problems more efficiently than classical methods. Many software engineering tasks, such as test case selection, static analysis, code clone detection, and defect prediction, involve complex optimization, search, or classification, making them candidates for quantum enhancement. In this paper, we introduce Quantum-Based Software Engineering (QBSE) as a new research direction for applying quantum computing to classical software engineering problems. We outline its scope, clarify its distinction from quantum software engineering (QSE), and identify key problem types that may benefit from quantum optimization, search, and learning techniques. We also summarize existing research efforts that remain fragmented. Finally, we outline a preliminary research agenda that may help guide the future development of QBSE, providing a structured and meaningful direction within software engineering.
\end{abstract}



\maketitle

\section{Introduction}

Many problems in software engineering involve optimization, search, or analysis in large, complex spaces~\cite{harman2012search}. Examples include selecting prioritized test cases for regression testing~\cite{yoo2012regression}, detecting code clones in large codebases~\cite{roy2009comparison}, and predicting defect-prone modules using historical data~\cite{hall2012systematic}. These tasks are computationally intensive, and classical solutions often rely on heuristics or approximation strategies~\cite{harman2001search, yoo2012regression}.

Quantum computing~\cite{nielsen2002quantum,preskill2018quantum} provides a fundamentally different computational model based on quantum mechanics. Quantum algorithms that exploit superposition, interference, and entanglement, such as Grover's search~\cite{grover1996fast} and the Quantum Approximate Optimization Algorithm (QAOA)~\cite{farhi2014quantum}, have shown advantages in domains such as combinatorial optimization. Quantum machine learning methods~\cite{biamonte2017quantum} also demonstrate potential in classification and pattern recognition tasks. These algorithmic features suggest that quantum-based approaches may help address certain software engineering problems, especially when classical methods are costly or ineffective~\cite{montanaro2016quantum, farhi2014quantum}.

In recent years, quantum computing research has focused primarily on quantum algorithm design~\cite{montanaro2016quantum}, hardware development~\cite{zhong2020quantum,arute2019quantum,preskill2018quantum}, and the development of quantum programming languages~\cite{heim2020quantum,gay2006quantum}. At the same time, Quantum Software Engineering (QSE) has emerged as a field that addresses the systematic development of quantum software, including specification, design, implementation, testing, and verification of programs intended to run on quantum hardware~\cite{zhao2020quantum,ali2022software,piattini2021toward,10.1145/3712002}. However, while building quantum software has attracted substantial attention, using quantum computation to support classical software engineering tasks remains relatively underexplored~\cite{wang2025quantum, mandal2025quantum,miranskyy2022quantum}. 

In this paper, we introduce Quantum-Based Software Engineering (QBSE) as a new research direction to explore how quantum computing can support classical software engineering tasks. Unlike QSE, which targets quantum software development, QBSE focuses on classical software engineering challenges and investigates whether quantum techniques can improve efficiency or scalability. We suggest that QBSE offers a timely and promising perspective that complements ongoing efforts in quantum software and classical engineering automation. In the following sections, we define the scope of QBSE, outline emerging applications, and present preliminary ideas for a research agenda.

\section{A Brief Background on Quantum Computing}
\label{sec:quantum background}

Quantum computing offers a fundamentally different model of computation, built on the principles of quantum mechanics~\cite{nielsen2002quantum}. This section introduces basic concepts that will help the reader understand the quantum techniques discussed later in this paper.


\begin{itemize}[leftmargin=2em] 
\setlength{\itemsep}{2.5pt}

\item \textit{Qubits.} Unlike classical bits that are either 0 or 1, a quantum bit (or qubit) can exist in a superposition of both. The state of a qubit, written as $|\varphi\rangle$, can be expressed as a combination of two basis states: $\alpha|0\rangle + \beta|1\rangle$, where $\alpha$ and $\beta$ are complex numbers such that $|\alpha|^2 + |\beta|^2 = 1$. The values $|\alpha|^2$ and $|\beta|^2$ represent the \textit{probability amplitudes}, indicating the likelihood of observing the qubit in state $|0\rangle$ or $|1\rangle$ upon measurement.

\item \textit{Quantum Gates and Circuits.} Quantum gates manipulate qubits through unitary operations. Common gates include the Hadamard gate (which creates superposition), the Pauli-X gate (analogous to a classical NOT), and the controlled-NOT (CNOT) gate, which is used to create entanglement. A quantum circuit consists of a sequence of such gates applied to one or more qubits.

\item \textit{Entanglement.} Entanglement is a quantum phenomenon in which the state of one qubit depends on the state of another, no matter how far apart they are. When two qubits are entangled, measuring one instantly determines the state of the other. This property is crucial for quantum algorithms that rely on coordination between qubits and is typically generated using gates like CNOT.

\item \textit{Measurement.} At the end of a quantum computation, the qubits are measured. This process collapses each qubit into a classical value, either 0 or 1, according to its probability amplitudes.

\item \textit{Quantum Algorithms.} Well-known quantum algorithms such as Grover’s search~\cite{grover1996fast} and Shor’s factoring~\cite{shor1999polynomial} use superposition, interference, and entanglement to achieve advantages over classical algorithms in specific problem domains.
\end{itemize}

For readers interested in a more in-depth treatment of these ideas, we refer to the comprehensive textbook by Nielsen and Chuang~\cite{nielsen2002quantum}.


\section{What is Quantum-Based Software Engineering?}

QBSE is a research direction that explores how quantum algorithms and hardware can be applied to solve problems in classical software engineering, such as test case selection, static analysis,
code clone detection, and defect prediction~\cite{pfleeger2009software}. It does not concern the development of quantum software itself.

QBSE is conceptually distinct QSE\footnote{In this paper, we use the terms QBSE and QSE for clarity and consistency. For convenience, these two directions can also be informally described as "quantum computing for software engineering" (QC4SE) and "software engineering for quantum computing" (SE4QC), respectively, following the conceptual distinction discussed in this paper.}
While QSE focuses on the engineering of quantum software systems, including design, programming languages, compilation, testing, and verification~\cite{zhao2020quantum}, QBSE explores how quantum computing can be used to enhance classical software engineering tasks.

Problems in classical software engineering that are combinatorial, search-based, or involve probabilistic reasoning are potential candidates for quantum support. Key tasks include testing, defect prediction, code clone detection, vulnerability analysis, and specification checking, all of which are discussed in more detail in Section~\ref{sec:applications}. These problems often involve large search spaces, uncertain heuristics, or high computational cost, making them suitable for exploration with quantum methods.

QBSE examines whether quantum techniques such as QAOA, Grover-based search, quantum annealing~\cite{kadowaki1998quantum}, or quantum machine learning can provide computational advantages in solving these tasks.

QBSE does not assume that all software engineering tasks are suitable for quantum computing. Instead, it focuses on identifying problem classes where quantum techniques may offer practical or theoretical benefits and developing methods to apply them effectively within software engineering workflows.

\section{Why We Need a New Research Direction}

Recent years have seen growing interest in applying quantum computing to classical software engineering tasks. While some promising early studies exist, they are often isolated and lack a shared framework, common terminology, or a unified research agenda. Defining QBSE as a distinct research direction can help address these gaps and guide future research efforts.

First, QBSE clarifies the scope by focusing on applying quantum computing to classical software engineering problems rather than developing software for quantum computers (as in QSE).

Second, having a defined research direction allows otherwise scattered efforts to be brought together under a common framework. This enables more systematic exploration and facilitates the development of shared benchmarks, problem formulations, and evaluation criteria.

Third, a recognized direction encourages the growth of a dedicated research community. It fosters collaboration, shared infrastructure, and the creation of venues for discussion, experimentation, and publication.

Finally, as quantum hardware and algorithms continue to improve, applying quantum computing to practical software engineering problems is becoming increasingly feasible. This makes it a good time to define QBSE and consider how it can guide future work in this research direction.


\section{Quantum Techniques}

This section introduces quantum computing techniques that may be useful in solving classical software engineering problems. These techniques are categorized into methods based on quantum search, optimization, machine learning, and annealing that have been explored or proposed in recent work. Table~\ref{tab:quantum-techniques-summary} summarizes the main classes of quantum techniques, along with representative algorithms and their hardware requirements. We then briefly review each category in the remainder of this section.
While this section focuses on the technical principles of quantum methods, we also briefly mention key software engineering applications for each technique. Detailed task-specific use cases are discussed in Section~\ref{sec:applications}.

\begin{table}[h]
\centering
\caption{Summary of Quantum Techniques Relevant to QBSE}
\footnotesize
\begin{tabular}{|p{3.5cm}|p{5cm}|p{4cm}|}
\hline
\textbf{Quantum Technique} & \textbf{Key Algorithms or Models} & \textbf{Hardware Requirements} \\
\hline
Quantum Search Algorithms & Grover's Algorithm~\cite{grover1996fast}, Amplitude Amplification~\cite{brassard2002quantum} & Gate-based Quantum Computers \\
\hline
Quantum Optimization Algorithms & QAOA~\cite{farhi2014quantum}, VQE~\cite{peruzzo2014variational}, QUBO Solvers~\cite{lucas2014ising} & Gate-based and Annealing Hardware \\
\hline
Quantum Machine Learning & QNN~\cite{beer2020training}, QSVM~\cite{rebentrost2014quantum}, Quantum Embeddings~\cite{schuld2019quantum} & Gate-based Quantum Computers or Simulators \\
\hline
Annealing-Based Optimization & D-Wave QUBO Solvers~\cite{johnson2011quantum}, Ising Model Formulations~\cite{kadowaki1998quantum} & Quantum Annealers (e.g., D-Wave) \\
\hline
\end{tabular}
\caption*{\scriptsize{\textit{Note:} The first three rows summarize algorithmic approaches, while the final row describes a hardware-specific optimization method based on quantum annealing.}}
\label{tab:quantum-techniques-summary}
\end{table}

\subsection{Quantum Search Algorithms}

Grover's algorithm~\cite{grover1996fast} is one of the most well-known quantum search techniques, offering a theoretical quadratic speedup for unstructured search problems. It has been formally analyzed for its optimality and extended into amplitude amplification methods~\cite{boyer1998tight, brassard2002quantum}.

Quantum search methods, such as Grover's algorithm, work by preparing a uniform superposition over all possible inputs and iteratively amplifying the amplitude of the target solution. A typical Grover iteration applies the operator:
$$
G = (2|\psi\rangle\langle\psi| - I)(I - 2|x_t\rangle\langle x_t|),
$$
where $|x_t\rangle$ is the desired solution and $|\psi\rangle$ is the initial uniform state. After $O(\sqrt{N})$ iterations, the probability of measuring the correct result is maximized.

In software engineering, quantum search algorithms have been explored in problems where large solution spaces are a bottleneck. Example scenarios include static analysis~\cite{ren2024dynamic} and finite-state machine (FSM) property checking~\cite{hall2009ses}. 

\subsection{Quantum Optimization Algorithms}

Quantum optimization techniques aim to address discrete optimization problems that are computationally intensive for classical approaches, especially those involving large or combinatorial search spaces. Among the most studied techniques is the Quantum Approximate Optimization Algorithm (QAOA)~\cite{farhi2014quantum}, a hybrid quantum-classical method designed for near-term quantum hardware. 

QAOA is particularly suited for problems formulated as Quadratic Unconstrained Binary Optimization (QUBO), a standard encoding for many NP-hard problems~\cite{lucas2014ising}.
A related formulation, Quadratic Unconstrained Directed Optimization (QUDO), extends QUBO by allowing directed dependencies among decision variables, enabling more expressive modeling in certain structured optimization problems. It alternates between cost and mixing Hamiltonians, expressed as:
$$
|\psi(\boldsymbol{\gamma}, \boldsymbol{\beta})\rangle = \prod_{j=1}^{p} e^{-i \beta_j H_M} e^{-i \gamma_j H_C} |\psi_0\rangle,
$$
where $|\psi_0\rangle$ is the initial state and $(\boldsymbol{\gamma}, \boldsymbol{\beta})$ are variational parameters to be optimized.

Another method, the Variational Quantum Eigensolver (VQE)~\cite{peruzzo2014variational}, was initially developed for quantum chemistry problems and has since been adapted for broader optimization tasks. Both QAOA and VQE are compatible with noisy intermediate-scale quantum (NISQ) devices~\cite{preskill2018quantum}, making them practical candidates for near-term experiments.

In software engineering, quantum optimization methods have been investigated for applications such as test case optimization~\cite{wang2024quantum,wang2023guess}. 

\subsection{Quantum Machine Learning}

Quantum machine learning (QML) explores how quantum computing might enhance learning models by leveraging quantum states and operations to increase expressiveness and computational capacity~\cite{biamonte2017quantum, schuld2015introduction}. A common approach involves using parameterized quantum circuits~\cite{schuld2020circuit} that encode classical input data in quantum states and apply trainable transformations.

One such model is the Quantum Neural Network (QNN), which uses layers of quantum gates to emulate the behavior of classical neural networks in a Hilbert space~\cite{beer2020training}. Another approach, the quantum support vector machine (QSVM), maps the classical input to a high-dimensional quantum feature space using circuit-based kernels~\cite{rebentrost2014quantum}.

A typical QNN applies a unitary transformation $U(\theta)$ to an input state $|\psi_0\rangle$, where $\theta$ represents trainable parameters optimized via classical feedback:
$$
|\psi(\theta)\rangle = U(\theta)|\psi_0\rangle
$$
The final state is then measured to compute output probabilities, which are used to evaluate loss functions and guide the training process.

In software engineering, QML methods have been explored for tasks such as defect prediction~\cite{nadim2024quantum} and vulnerability detection~\cite{zhou2022new}.

\subsection{Annealing-Based Optimization}

Annealing-based optimization refers to a class of quantum techniques that use adiabatic quantum evolution to solve combinatorial optimization problems~\cite{kadowaki1998quantum, albash2018adiabatic}. Unlike gate-based quantum algorithms (e.g., QAOA), these approaches are typically implemented on quantum annealing hardware such as D-Wave~\cite{johnson2011quantum}, which encodes binary variables using physical qubits and evolves the system toward a low-energy state corresponding to an approximate solution.

A typical formulation involves minimizing an Ising Hamiltonian:
\[
H = \sum_i h_i \sigma_i^z + \sum_{i<j} J_{ij} \sigma_i^z \sigma_j^z,
\]
where $\sigma_i^z$ denotes the Pauli-Z operator on qubit $i$, $h_i$ are local bias terms, and $J_{ij}$ represent coupling strengths between qubits. The system starts in a ground state of a simple initial Hamiltonian and evolves toward the problem Hamiltonian over time.

In software engineering, annealing-based methods have been explored for test case generation~\cite{araujo2025using}, test suite minimization~\cite{wang2024test}, and regression test optimization~\cite{trovato2025reformulating}.

\vspace*{2mm}
Together, these techniques offer promising tools for addressing computationally intensive problems in classical software engineering. As quantum hardware and tools mature, it may become increasingly feasible to integrate such approaches into practical development workflows.
While these techniques show early promise, many remain at the conceptual or prototype level, and their integration into practical software engineering workflows is still evolving.

\begin{table}[t]
\footnotesize
\centering
\caption{Mapping Quantum Techniques to Classical Software Engineering Tasks}
\label{tab:qse_summary}
\begin{tabular}{|p{4.5cm}|p{5.5cm}|p{2.6cm}|}
\hline
\textbf{Software Engineering Tasks} & \textbf{Quantum Techniques Used} & \textbf{Example References} \\
\hline
Test Case Optimization & QAOA, Quantum Annealing & \cite{wang2024quantum, wang2023guess} \\
\hline
Test Suite Minimization & Grover’s Algorithm, Quantum Annealing & \cite{hussein2021quantum,wang2025bqtmizer,wang2024test} \\
\hline
Test Case Generation & Grover’s Algorithm, Quantum Annealing & \cite{hall2009ses,araujo2025using} \\
\hline
Regression Testing & Grover’s Algorithm, Quantum Annealing, Quantum Counting, QAOA & \cite{miranskyy2022using, trovato2025reformulating,trovato2025preliminary} \\
\hline
System-Level Testing & QNN & \cite{wang2024quantum-1} \\
\hline
Software Vulnerability Detection & QNN & \cite{zhou2022new, song2024recurrent} \\
\hline
Supply Chain Attack and Vulnerability Detection & QSVM, QNN & \cite{masum2022quantum, akter2022software} \\
\hline
Defect Prediction & QSVM, VQC, Quantum Annealing & \cite{nadim2024quantum, nadim2025comparative,mandal2024evaluating} \\
\hline
Static Analysis & Grover’s Algorithm & \cite{ren2024dynamic} \\
\hline
Code Clone Detection & Quantum Annealing (QUBO/QUDO) & \cite{jhaveri2023cloning} \\
\hline
Finite State Machine Property Checking & Grover's Algorithm & \cite{hall2009ses} \\
\hline
Component Synthesis from Library & Grover's Algorithm & \cite{hall2009ses} \\
\hline
QoS Prediction & QELM & \cite{wang2024application} \\
\hline
\end{tabular}
\end{table}

\section{Emerging Applications in Quantum-Based Software Engineering}
\label{sec:applications}

In this section, we present classical software engineering tasks that may benefit from quantum computing techniques. These tasks often involve large search spaces, combinatorial complexity, or high-dimensional data representations, making them suitable candidates for quantum optimization, search, and learning methods.
We focus on organizing emerging studies that concretely demonstrate how quantum computing can be applied to classical software engineering problems. We categorize the literature by task, including testing, defect prediction, code analysis, vulnerability detection, and specification checking, and highlight the quantum techniques explored in each case. While these efforts remain exploratory, they form an initial base for QBSE.



\subsection{Test Case Optimization and Minimization}

Selecting a minimal or high-priority subset of test cases that satisfies given coverage criteria is a well-known combinatorial problem~\cite{yoo2012regression}. Classical heuristics often struggle with scalability as test suites grow. Quantum optimization techniques such as QAOA and quantum annealing have been proposed as alternatives to explore large solution spaces more efficiently.

Several studies have applied these methods to optimize test cases~\cite{wang2024quantum,wang2023guess}. QUBO formulations and Grover's algorithm have also been used for test suite minimization~\cite{hussein2021quantum, wang2024test}. In addition, quantum-enhanced search has been explored to generate test cases~\cite{hall2009ses}, and Grover-based search and quantum counting~\cite{boyer1998tight} have been used to accelerate dynamic regression testing~\cite{miranskyy2022using}.

\subsection{Defect and QoS Prediction}

Defect prediction aims to identify fault-prone modules using historical data, software metrics, or graph representations~\cite{lessmann2008benchmarking}. Quantum machine learning models such as QNNs and QSVMs are promising tools for this task, particularly in small or unbalanced datasets where quantum encodings may help improve generalization~\cite{beer2020training, rebentrost2014quantum}.

Recent studies suggest that QSVMs may outperform classical SVMs in certain software defect datasets~\cite{nadim2024quantum}. Additional experiments show that QNNs may require fewer resources than QSVMs in certain settings~\cite{nadim2025comparative}, highlighting differences in model performance that depend on the architecture and data regime.

Quantum machine learning has also been applied to performance prediction tasks. For example, Wang et al.~\cite{wang2024application} utilized Quantum Extreme Learning Machines (QELM)~\cite{mujal2021opportunities} to estimate system-level quality attributes, such as response time, in industrial control systems, demonstrating the potential of quantum models beyond defect prediction.

\subsection{Code Clone Detection and Static Analysis}

Code clone detection is important for software maintenance and refactoring~\cite{roy2009comparison}. It involves identifying syntactic or semantic similarities among code fragments. Classical techniques, especially those using tokens, AST, or graph-based comparison, are computationally expensive on large codebases. Quantum annealing techniques such as QUBO and QUDO have been proposed to reduce detection cost~\cite{jhaveri2023cloning}.

Similarly, static analysis tasks such as reachability and transitive closure computation can be enhanced by Grover’s algorithm~\cite{ren2024dynamic}, which offers quadratic speedups in certain search-based formulations.

\subsection{Vulnerability Detection and Software Security}

Quantum-enhanced models have been investigated for software vulnerability detection, especially in source code embeddings and token classification tasks. QNNs, QSVMs, and quantum embedding models~\cite{beer2020training, schuld2019quantum} have been used to classify vulnerable patterns. For example, Zhou et al.~\cite{zhou2022new} trained a QNN on tokenized code, and Song et al.~\cite{song2024recurrent} proposed a recurrent quantum embedded neural network (RQENN) to model context-sensitive patterns.

In addition, quantum classifiers have been applied to secure software supply chains. Masum et al.~\cite{masum2022quantum} focused on detecting software supply chain attacks by analyzing metadata and dependency features using quantum machine learning models. Akter et al.~\cite{akter2022software}, on the other hand, addressed supply chain vulnerability detection and showed that quantum models outperform classical baselines in small-sample settings.

\subsection{Specification Checking and Component Synthesis}

Formal verification tasks, such as property checking, can benefit from a quantum search. Grover-based algorithms have been applied to reachability checking in finite-state machines~\cite{hall2009ses}, including variations that handle unknown target counts. Quantum-enhanced search has also been proposed to synthesize components from libraries based on input-output constraints~\cite{hall2009ses}, although these approaches remain in the proof-of-concept stage and have not yet been validated in realistic settings.

\subsection{Positioning and Summary of Emerging QBSE Applications}

Table~\ref{tab:qse_summary} summarizes the software engineering tasks discussed above, the quantum techniques applied to them, and illustrative references. While many applications remain in the early stages, the reviewed work suggests that QBSE is a promising direction with growing interest and early experimental support. 

In addition to these task-specific studies, several broader efforts have attempted to map quantum algorithm classes to software engineering challenges. Miranskyy et al.~\cite{miranskyy2022quantum} outlined how eight classes of quantum algorithms, including Grover's search, quantum SAT solvers, quantum linear systems, and quantum walks, could be applied to various phases of software development. Their work offers a conceptual roadmap, but does not examine specific implementations. 
Mandal et al.~\cite{mandal2025quantum} reviewed existing research applying quantum computing to software engineering and categorized it into areas such as optimization, search, and machine learning. Their paper emphasized potential integration workflows, but did not systematically organize empirical studies by task type.
Wang et al.~\cite{wang2025quantum} focused on the role of Quantum Artificial Intelligence (QAI), particularly quantum optimization and quantum machine learning, in software engineering. Their paper outlined challenges such as algorithm design, noise handling, and problem representation, and suggested potential research directions. While this work offers valuable insights into the role of quantum AI within software engineering, its focus represents only one aspect of the broader QBSE landscape. 

Compared with these works, our paper aims to help establish QBSE as a distinct research direction by systematically organizing and interpreting early applications of quantum computing to classical software engineering tasks. By highlighting the tasks, techniques, and emerging evidence in a structured manner, we seek to define the scope and significance of QBSE within the broader software engineering landscape.

\section{A Preliminary Research Agenda}

QBSE is still in its early stages of development. To support its development, we outline a preliminary agenda of near-term directions.

\begin{itemize}[leftmargin=2em] 
\setlength{\itemsep}{2.5pt}
\item \textit{Problem reformulation and suitability analysis.} Not all software engineering problems are suitable for quantum methods. A good starting point is to identify problems with features such as combinatorial complexity, discrete or combinatorial search spaces, or high-dimensional input. Such problems can sometimes be reformulated into quantum-friendly representations, including QUBO~\cite{lucas2014ising}, Ising models~\cite{farhi2014quantum}, or parameterized quantum circuits (PQCs)~\cite{peruzzo2014variational}.

\item \textit{Design of quantum-assisted methods.} Once suitable formulations are in place, the next step is to develop quantum-assisted methods that integrate quantum algorithms into classical software engineering workflows. These may include quantum routines for tasks such as test case selection, defect prediction, or code clone detection, potentially embedded within larger software engineering pipelines, although such integrations remain at an early stage.

\item \textit{Empirical evaluation and benchmarking.} Empirical evaluation is essential to assess the value of QBSE approaches. This involves defining benchmark tasks, selecting meaningful baselines, and measuring performance on both quantum hardware and simulators. Both solution quality and scalability should be considered.

\item \textit{Tool and framework development.} Prototype tools and reusable components are important for reproducibility and adoption. Examples include wrappers for existing tools, libraries for encoding quantum models, and standard interfaces for hybrid execution environments.

\item \textit{Cross-disciplinary collaboration.} QBSE sits at the intersection of quantum computing and software engineering, so interdisciplinary collaboration is key. Research teams should combine their expertise to ensure that quantum techniques are applied effectively while keeping software engineering goals at the forefront.
\end{itemize}

This agenda is intended to guide early work on QBSE. As the field progresses, more specific research topics and technical challenges will naturally emerge.

\section{Conclusion}

In this paper, we introduced Quantum-Based Software Engineering (QBSE) as a new research direction. QBSE focuses on applying quantum computing techniques to solve classical software engineering problems. It differs from QSE, which focuses on developing software for quantum systems.

We outlined the scope of QBSE, identified key software engineering tasks that may benefit from quantum computation, and summarized relevant quantum techniques. We also reviewed existing work that aligns with the QBSE perspective, but remains fragmented. To support further exploration of this direction, we proposed a preliminary research agenda covering problem reformulation, method design, empirical evaluation, tool support, and collaboration.

By establishing a clearer perspective and structure, QBSE may offer a foundation to organize existing research and guide future efforts. We hope that this work encourages further exploration and cooperation between the software engineering and quantum computing communities. We also invite researchers to contribute further to the refinement of QBSE by proposing new use cases, frameworks, and validation methods as the field continues to evolve.

\bibliographystyle{acm}
\bibliography{ref}

\end{document}